# Composition of Near-Earth Asteroid (4179) Toutatis

Vishnu Reddy[1]
Department of Space Studies, University of North Dakota, Grand Forks, USA.
Max Planck Institute for Solar System Research, Katlenburg-Lindau, Germany.
Email: reddy@space.edu

Juan A. Sanchez
Institut für Planetologie, University of Münster, 48149 Münster, Germany.

Michael J. Gaffey[1]
Department of Space Studies, University of North Dakota, Grand Forks, ND 58202, USA.

Paul A. Abell[1]
Astromaterials Research & Exploration Science Directorate, NASA Johnson Space Center, Mail Code KR, 2101 NASA Parkway, Houston, TX 77058-3696, USA.

Lucille Le Corre[1]
Max Planck Institute for Solar System Research, Katlenburg-Lindau, Germany 37191

Paul S. Hardersen[1]
Department of Space Studies, University of North Dakota, Grand Forks, ND 58202, USA.

Pages: 15
Figures: 2
Tables: 1

**Proposed Running Head:** Composition of Toutatis

**Editorial correspondence to:**
Vishnu Reddy
Max-Planck Institute for Solar System Research,
37191 Katlenburg-Lindau,
Germany.
 +49-5556-979-550, reddy@mps.mpg.de

**Abstract**

Surface composition of near-Earth asteroid (4179) Toutatis is consistent with an undifferentiated L-chondrite composition. This is inconsistent with early observations that suggested high pyroxene iron content and a differentiated body.

# 1. Introduction

Near-Earth asteroid (NEA) (4179) Toutatis is an Apollo asteroid that has been the focus of several ground-based observational campaigns using different remote sensing techniques (photometry, spectroscopy and radar). One of the first studies of Toutatis was conducted by Spencer et al. (1995), who generated phase curves using the IAU H, G system developed by Bowell et al. (1989). From these photometric data, they obtained a mean H of 15.3 and a slope parameter G of 0.10 for the asteroid. From Toutatis' lightcurve, Spencer et al. (1995) determined that the asteroid has a complex rotation with a period between 3 and 7.3 days. Hudson and Ostro (1995), used radar observations to reconstruct Toutatis' shape and spin state. They found that Toutatis has an irregular two-lobes shape, whose dimensions along the principal axes are 1.92, 2.40 and 4.60 km. This study showed that Toutatis has a non-principal axis rotation. The rotation period around its long axis ($P\psi$) was found to be 5.41 days, with an average of 7.35 days for the long-axis precession period ($P\phi$). These numbers were slightly refined by Ostro et al. (1999), who found values of $P\psi$=5.37 and $P\phi$=7.42. These results were later confirmed by the findings of Mueller et al. (2002).

Hudson and Ostro (1998) combined the data from Spencer et al. (1995), and Hudson and Ostro (1995) to estimate the Hapke parameters (see e.g., Hapke 1981, Hapke 1993) of Toutatis. They found a particle single-scattering albedo $w = 0.261\pm 0.019$, opposition surge width $h = 0.036\pm0.023$, a particle phase function asymmetry factor $g = -0.29\pm0.06$, an amplitude $B_0$=1.20±0.32, and macroscopic roughness parameter $\theta = 32\pm8°$. The analysis of Hudson and Ostro (1998) indicated that a fine regolith layer covers a large part of the surface of Toutatis. Howell et al. (1994), obtained spectroscopic and spectrophotometric data (0.5-3.35 μm) of Toutatis during its opposition on January 1993. They found that the spectrum of Toutatis exhibit a pyroxene absorption band centered at 1.966 μm. The VIS spectrum of Toutatis indicated that the Band I center is located at wavelengths longer than 0.96 μm, suggesting the presence of some olivine. Based on the work of Adams (1974), and from the measured band centers of Toutatis, Howell et al. (1994) determined that its average pyroxene composition was $Fs_{45-55}En_{50-35}Wo_{5-10}$. After applying the thermophysical model to the 3-μm spectral region of Toutatis, Howell et al. (1994) found a thermal inertia of $3-8 \times 10^5$ erg cm$^{-2}$ K$^{-1}$ sec$^{-1/2}$, a geometrical albedo ≥0.20, a beaming parameter ≥ 0.9, an emissivity ≥ 0.9, and a phase coefficient in the range of 0.005-0.01.

Additional spectral data of Toutatis were obtained by Davies et al. (2007), who acquired NIR spectra (1.0-2.5 μm) with the UKIRT telescope over a wide range of phase angles (0.7°-81°). These observations were intended to seek evidence for phase reddening and to search for possible surface variations with rotational phase. However, within the uncertainties of their data neither phase reddening nor surface heterogeneities were detected. In this work we present detailed compositional analysis of Toutatis and constrain its surface mineralogy. Our interest in this object is heightened by the possibility of a flyby by the Chinese lunar spacecraft Chang'e 2 on December 13, 2012 (William Gary, personal comm). Although no official conformation of this event is available, optical astrometry of the Chang'e 2 spacecraft strongly support such an event.

**2. Observations and data reduction**

Observations of (4179) Toutatis were obtained remotely with the SpeX instrument (Rayner et al., 2003) on the NASA IRTF, Mauna Kea, Hawai'i. The NIR spectra (~0.7-2.5 μm) were acquired using SpeX in low-resolution (R~150) prism mode with a 0.8" slit width. The asteroid was observed on August 13, 2008 when its V-mag was 14.8. To correct for telluric water vapor features, and to obtain the relative reflectance, standard and solar analog stars were observed at similar air masses as the asteroid. Flat fields and lamp spectra were also acquired. SpeX prism data were processed using the IDL-based Spextool provided by the NASA IRTF (Cushing et al. 2004). Detailed description of data reduction procedure is available in Reddy et al. (2012).

**3. Analysis**

The near-IR spectrum of Toutatis is shown in Fig. 1 (solid circles). This spectrum exhibits two absorption bands characteristics of olivine-orthopyroxene assemblages, and it is consistent with previous observations of this asteroid (e.g., Davies et al. 2007, Dunn and Burbine 2012). In order to extent the wavelength range we combined our NIR spectrum of Toutatis with its visible spectrum (Fig 1, open triangles) obtained by the SMASS survey (Bus and Binzel, 2002a,b). Spectral band parameters, band centers, band depths, and band area ratio (BAR) were measured in the same way as in Sanchez et al. (2012). BAR is the ratio of area of Band II to that of Band I. Temperature corrections to Band II center; Band II depth and BAR were applied using the equations from Sanchez et al. (2012). No corrections were applied for Band I center. Olivine and pyroxene chemistries were determined using the spectral calibrations derived by Dunn et al. (2010a). Toutatis has a Band I center of 0.94±0.01 μm, a Band II center of 1.98±0.03 μm and a BAR of 0.5±0.05. The pyroxene absorption bands themselves have band depths of 15±0.3% (Band I) and 4±0.5% (Band II). The Band I center of Toutatis is longer than what is normally expected for pure low calcium pyroxene (0.90-0.93 μm). This would suggest the presence of a second pyroxene phase like high calcium pyroxene (Band I center range 0.91-1.06 μm). The Band II center of Toutatis (1.98±0.03 μm) is consistent with those of low calcium (1.8-2.1 μm) and high calcium pyroxenes (1.97-2.35 μm). Spectral band parameters and mineral chemistry values are presented in Table 1. A side lobe at 1.3 μm suggests the possible presence of olivine, which can also affect the Band I center and cause it to shift to longer wavelengths.

Figure 2A shows a close up of the S(IV) region of the Gaffey S-asteroid subtypes. The measured Band I center and BAR of Toutatis are depicted as a filled triangle. Based on these band parameters, Toutatis is classified as S(IV) in the system of Gaffey et al. (1993), which is consistent with an ordinary chondrite-like composition for this asteroid. The dotted line shows the olivine-orthopyroxene (low calcium pyroxene) mixing line and Toutatis band parameters are offset from it to the left, which is an indication for the presence of high calcium pyroxene in the surface. The asteroid plots at the boundary between H and L chondrites within the S(IV) region.

Pyroxene and olivine abundance and chemistry can be used to constrain the surface composition and identify meteorite analogs. Using equations from Dunn et al. (2010a) we calculate the olivine and pyroxene chemistry to be $Fa_{20\pm1.3}(Fo_{80\pm1.3})$ and $Fs_{17\pm1.4}$, respectively. The ratio of olivine in a mixture of olivine and pyroxene is 0.61±0.03. These values are within the olivine and pyroxene chemistry ranges derived by Dunn et al. (2010) for H ($Fa_{15-21}$ and $Fs_{13-19}$) and L ($Fa_{21-27}$ and $Fs_{17-23}$) chondrites. Independent olivine chemistry ($Fa_{22-26}$) and abundance (0.53-0.66) estimates by Dunn and Burbine (2012) using a different set of observations are also consistent with our values ($Fa_{20\pm1.3}$ and 0.61±0.03). However, their values are more consistent with an L/LL chondrite type assemblage compared to H/L type from our data. This discrepancy could be attributed to the use of a different technique when calculating the band parameters. Combining Dunn and Burbine (2012) with our olivine and pyroxene chemistry and abundance values we get mean values of $Fa_{23\pm1.3}$, $Fs_{19.5\pm1.4}$, and 0.60±0.03, respectively. These values are consistent with an L chondrite type surface assemblage, which we think is the best analog for Toutatis' surface.

However, pyroxene chemistry values ($Fs_{45-55}$) estimated by Howell et al. (1994) are inconsistent with our values ($Fs_{19.5\pm1.4}$). Based on this high ferrosilite content ($Fs_{45-55}$), Howell et al. (1994) suggested Toutatis parent body "had considerable differentiation of the silicates." We believe these differences are primarily due to incomplete spectral coverage of the Band I and II features in their data, unavailability of thermal correction equations of band parameters (Sanchez et al. 2012), and the lack of robust laboratory calibration for determining mineral chemistry (Dunn et al. 2010a) at that time. Figure 2B shows the molar contents of Fa vs. Fs for Toutatis (filled square), along with the values for LL (open triangles), L (open circles) and H (x symbols) ordinary chondrites from Nakamura et al. (2011). As with Fig. 2A, Toutatis plots in the boundary region between H and L chondrites consistent with olivine and pyroxene chemistries and abundance calculated from our observations.

**4. Discussion**

Surface composition of Toutatis is consistent with an L chondrite type assemblage based on the mean olivine and pyroxene chemistry and the abundance of olivine. Toutatis' olivine and pyroxene chemistries are similar to those of (433) Eros ($Fa_{20}$ and $Fs_{24}$), the target of the NEAR mission (Izenberg et al. 2003). However, the olivine abundance is higher on Eros (0.65-0.69) compared to Toutatis (0.60). Typically increasing olivine abundance is a sign of increasing oxidation where L and LL chondrites (which are more oxidized) have higher olivine abundance (0.52-0.72) than H chondrites (0.44-0.62) (Dunn et al. 2010b). Oxygen isotope analysis also shows that H, L and LL chondrites formed on different parent bodies (Clayton, 1993).

Nesvorny et al. (2009) suggested the Gefion family is the source of L chondrites based on dynamical modeling. The formation of this family is also linked to the influx of L chondrite meteorites to the Earth as evident from fossil meteorites discovered in limestone quarry in southern Sweden that all date ~467 million years. A majority of the L chondrites in our terrestrial meteorite collection are heavily shocked and have $^{39}$Ar-$^{40}$Ar age near 470 million years (Korochantseva et al., 2007). This suggests a major impact event on the L chondrite parent body ~470 million years ago that led to the formation of the Gefion family (Nesvorny et al. 2009). If Toutatis was formed as part of the Gefion family, it was transported to the 5:2 resonance with Jupiter (Nesvorny et al. 2009) via Yarkovsky effect and then to near-Earth space more recently (few million years). This is because the collisional lifetime of the NEA is much shorter (few tens of millions of years) than the age of the Gefion family (470 million years) (Bottke et al. 2005). Cosmic ray exposure (CRE) ages of L chondrites suggest that they were spalled off their parent body (in the main belt or near-Earth asteroid population) on two occasions (~28 and ~40 million years). It is possible that some L chondrites in our terrestrial collection were spalled off Toutatis as meter-sized objects during those events, although these events are at the limits of NEA lifetimes in the inner solar system.

182    Ostro et al. (1999) suggested that 1/3$^{rd}$ of Toutatis' surface was covered by a
183 'smooth component' or regolith with porosity of lunar soils. They argued that if the
184 surface composition of Toutatis was like an ordinary chondrite, then this smooth
185 component is a solid layer with no "more than a centimeter of overlaying regolith" (Ostro
186 et al. 1999). The shape derived from radar observations, along with its surface regolith
187 properties and mineralogy suggests that Toutatis' surface would be similar to that of
188 (25143) Itokawa, which was studied by the Hayabusa spacecraft. Our observations and
189 predictions about Toutatis' surface properties would be confirmed if the Chang'e-2
190 spacecraft makes a successful flyby of Toutatis.
191
192 **Acknowledgement**

193 This research was supported by NASA NEOO Program Grant NNX12AG12G, and
194 NASA Planetary Geology and Geophysics Grant NNX11AN84G. We thank the IRTF
195 TAC for awarding time to this project, and to the IRTF TOs and MKSS staff for their
196 support.

197

198 **References**

199 Adams, J.B., 1974. Visible and near infrared diffuse reflectance spectra of pyroxenes as
200 applied to remote sensing of solid objects in the solar system. Journal of Geophysical
201 Research 79, 4829-4836.
202
203 Bottke, W.F., Durda, D., Nesvorný, D., Jedicke, R., Morbidelli, A., Vokrouhlický, D.,
204 Levison, H.F., 2005. Linking the collisional history of the main asteroid belt to its dynamical excitation and depletion. Icarus 179, 63–94.

Bowell, E., Hapke, B., Domingue, D., Lumme, K., Peltoniemi, J., Harris, A.W., 1989. Application of photometric models to asteroids. In: Binzel, R.P., Gehrels, T., Matthews, M.S. (Eds.), Asteroids II. Univ. Arizona Press, pp. 524-556.

Bus, S.J., Binzel, R.P., 2002a. Phase II of the small main-belt asteroid spectroscopic survey a feature-based taxonomy. Icarus 158, 146-177.

Bus, S.J., Binzel, R.P., 2002b. Phase II of the small main-belt asteroid spectroscopic survey the observations. Icarus 158, 106-145.

Clayton, R.N., 1993. Oxygen isotopes in meteorites. Annu. Rev. Earth Planet. Sci. 21, 115–149.

Cloutis, E.A., Gaffey, M.J., Jackowski, T.L., Reed, K.L., 1986. Calibrations of phase abundance, composition, and particle size distribution for olivine- orthopyroxene mixtures from reflectance spectra. Journal of Geophysical Research 91, 11641-11653.

Cushing, M.C., Vacca, W.D., Rayner, J.T., 2004. Spextool: A spectral extraction package for SpeX, a 0.8–5.5 micron cross-dispersed spectrograph. Publ. Astron. Soc. Pacific 116 (818), 362–376.

Davies, John K., Harris, Alan W., Rivkin, Andrew S., Wolters, Stephen D., Green, Simon F., McBride, Neil, Mann, Rita K., Kerr, Tom H., 2007. Near-infrared spectra of 12 Near-

Earth Objects. Icarus 186, 111-125.

Dunn, T.L., McCoy, T.J., Sunshine, J.M., McSween, H.Y., 2010a. A coordinated spectral, mineralogical, and compositional study of ordinary chondrites. Icarus 208, 789-797.

Dunn, T.L., McSween, Jr., H.Y.., McCoy, T.J., Cressey, G., 2010b. Analysis of ordinary chondrites using powder X-ray diffraction: 2. Applications to ordinary chondrite parent-body processes. Meteoritics and Planetary Science 45, 135-156.

Dunn, T. L., and Burbine, T. H. 2012. Mineralogies of Near Earth Asteroids, 43rd Lunar and Planetary Science Conference, held March 19–23, 2012 at The Woodlands, Texas. LPI Contribution No. 1659, id.2305

Gaffey, M.J., Burbine, T.H., Piatek, J.L., Reed, K.L., Chaky, D.A., Bell, J.F., Brown, R.H., 1993. Mineralogical variations within the S-type asteroid class. Icarus 106, 573-602.

Hapke, B., 1993. Theory of Reflectance and Emittance Spectroscopy. Cambridge University Press, Cambridge, UK.

Hapke, B., 1981. Bidirectional reflectance spectroscopy: I - Theory. J. Geophys. Res. 86, 3039–3054.

Howell, E.S., Britt, D.T., Bell, J.F., Binzel, R.P., Lebofsky, L.A., 1994. Visible and near-infrared spectral observations of 4179 Toutatis. Icarus 111, 468-474.

Hudson, R. S., and Ostro, S. J., 1998. Photometric properties of Asteroid 4179 Toutatis from lightcurves and a radar-derived physical model. Icarus 135 451-457.

Hudson, R. S., and Ostro, S. J., 1995. Shape and non-principal-axis spin state of asteroid 4179 Toutatis. Science 270, 84-86.

Izenberg, N.R., Murchie, S.L., Bell III, J.F., McFadden, L.A., Wellnitz, D.D., Clark, B.E., Gaffey, M.J., 2003. Spectral properties and geologic processes on Eros from combined NEAR NIS and MSI data sets. Meteorit. Planet. Sci. 38, 1053–1077.

Korochantseva, E.V., Trieloff, M., Lorenz, C.A., Buykin, A.I., Ivanova, M.A., Schwarz, W.H., Hopp, J., Jessberger, E.K., 2007. L-chondrite asteroid breakup tied to Ordovician meteorite shower by multiple isochron 40Ar–39Ar dating. Meteorit. Planet. Sci. 42, 113–130.

Mueller, Béatrice E. A., Samarasinha, Nalin H., Belton, Michael J. S., 2002. The Diagnosis of Complex Rotation in the Lightcurve of 4179 Toutatis and Potential Applications to Other Asteroids and Bare Cometary Nuclei. Icarus 158, 305-311.

Nakamura, T., Noguchi, T., Tanaka, M., Zolensky, M.E., Kimura, M., Tsuchiyama, A., Nakato, A., Ogami, T., Ishida, H., Uesugi, M., Yada, T., Shirai, K., Fujimura, A., Okazaki, R., Sandford, S.A., Ishibashi, Y., Abe, M., Okada, T., Ueno, M., Mukai, T., Yoshikawa, M., Kawaguchi, J., 2011. Itokawa Dust Particles: A Direct Link Between S-Type Asteroids and Ordinary Chondrites. Science 333, 1113-1116.

Nesvorný, D., Vokrouhlický, D., Morbidelli, A., Bottke, W. F. 2009. Asteroidal source of L chondrite meteorites, Icarus, Volume 200, Issue 2, p. 698-701.

Ostro, S.J., Hudson, R.S., Rosema, K.D., Giorgini, J.D., Jurgens, R.F., Yeo-mans, D.K., Chodas, P.W., Winkler, R., Rose, R., Choate, D., Cormier, R.A., Kelley, D., Littlefair, R., Benner, L.A.M., Thomas, M.L., Slade, M.A., 1999. Asteroid 4179 Toutatis: 1996 Radar Observations. Icarus 137, 122-139.

Rayner, J.T., Toomey, D.W., Onaka, P.M., Denault, A.J., Stahlberger, W.E., Vacca, W.D., Cushing, M.C., Wang, S., 2003. SpeX: A Medium-Resolution 0.8-5.5 Micron Spectrograph and Imager for the NASA Infrared Telescope Facility. Publications of the Astronomical Society of the Pacific 115, 362-382.

Reddy, V, Sanchez, J. A., Nathues, A., Moskovitz, N. A., Li, J-Y., Cloutis, E. A., Archer, K., Tucker, R. A., Gaffey, M. J., Mann, P. J., Sierks, H., Schade, U., 2012. Photometric, spectral phase and temperature effects on 4 Vesta and HED meteorites: Implications for the Dawn mission, Icarus, 217:153-168.

Sanchez, J.A., Reddy, V., Nathues, A., Cloutis, E.A., Mann, P., Hiesinger, H., 2012. Phase reddening on near-Earth asteroids: Implications for mineralogical analysis, space weathering and taxonomic classification. Icarus 220, 36-50.

Spencer, J.R., et al., 1995. The lightcurve of 4179 Toutatis: Evidence for complex rotation. Icarus 117, 71-89.

**Table 1. Spectral band parameters, ol/(ol+px) ratio, and olivine and pyroxene chemistry for 4179 Toutatis.**

| Band I Center (μm) | Band II Center (μm) | Band I Depth (%) | Band II Depth (%) | Band Area Ratio | olivine (olv+pyx) | Fayalite (mol %) | Ferrosilite (mol %) |
|---|---|---|---|---|---|---|---|
| 0.94±0.01 | 1.98±0.03 | 15±0.3 | 4±0.5 | 0.5±0.05 | 0.61±0.03 | 20.2±1.3 | 17.4±1.4 |

**Figure Captions**

Figure 1. NIR spectrum of 4179 Toutatis (solid circles) obtained with the SpeX instrument on NASA IRTF. The spectrum exhibits the two absorption bands characteristic of olivine-pyroxene assemblages. The visible spectrum of Toutatis (open triangles) was taken from the SMASS survey (Bus and Binzel, 2002a,b).

Figure 2. (a) Plot of the Band I center versus BAR for 4179 Toutatis (filled triangle). The polygonal region corresponding to the S(IV) subgroup represents the mafic silicate

components of ordinary chondrites (Gaffey et al., 1993). The dashed curve indicates the location of the olivine-orthopyroxene mixing line (Cloutis et al., 1986). The horizontal lines represent the approximate boundaries for LL, L, and H ordinary chondrites found by Dunn et al. (2010a). (b) Molar contents of Fa vs. Fs for 4179 Toutatis (filled square) calculated using the spectral calibrations of Dunn et al. (2010a). For comparison, values for LL (open triangles), L (open circles) and H (x) ordinary chondrites from Nakamura et al. (2011) are also shown. The horizontal dashed-lines represent the approximate boundaries for LL, L, and H ordinary chondrites.

**Figure 1. Composition of Toutatis**

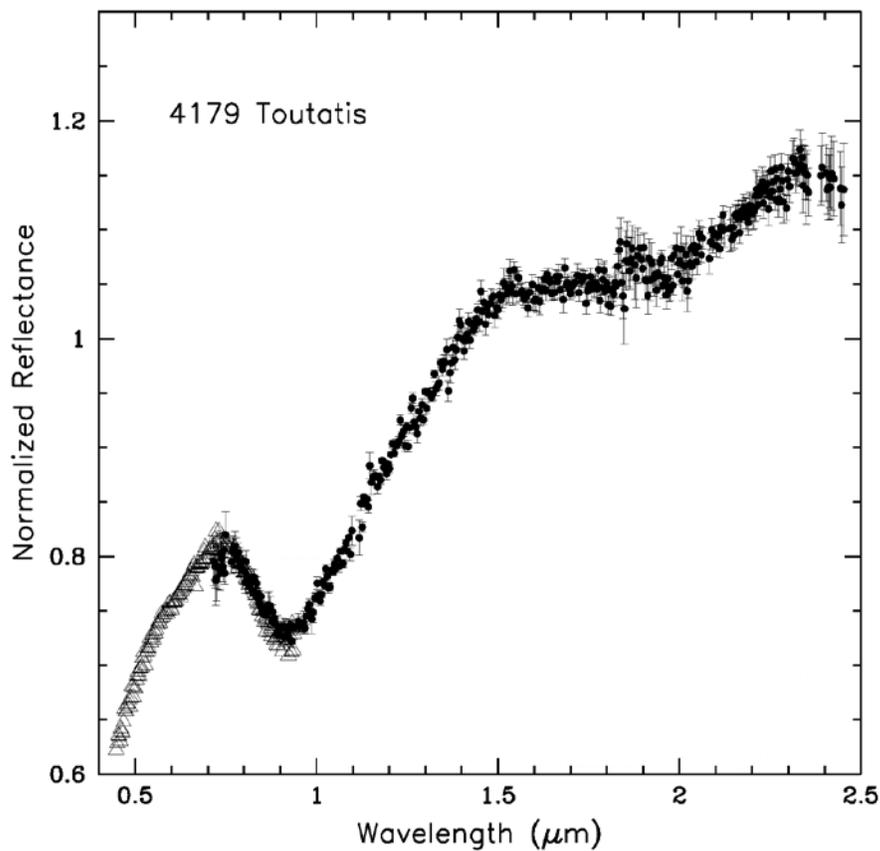

331    **Figure 2. Composition of Toutatis**

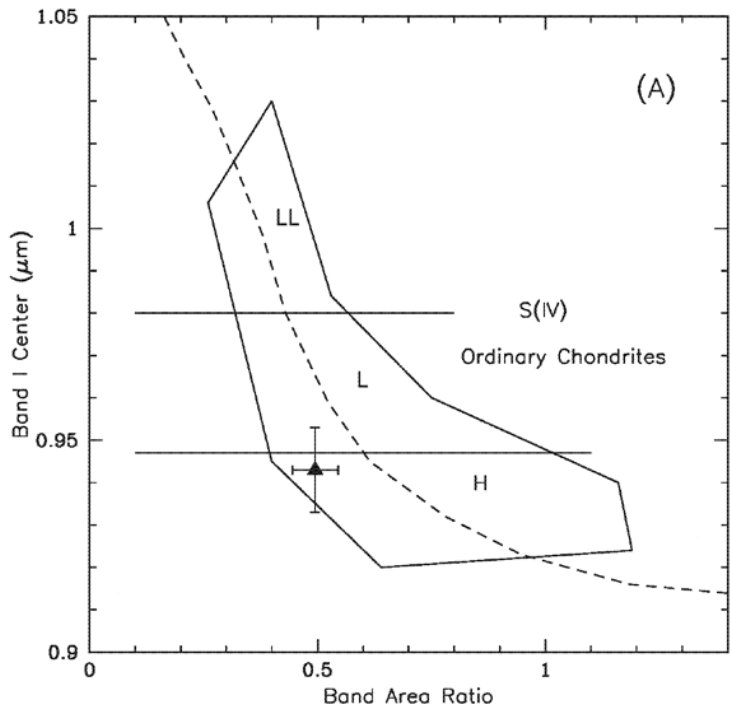
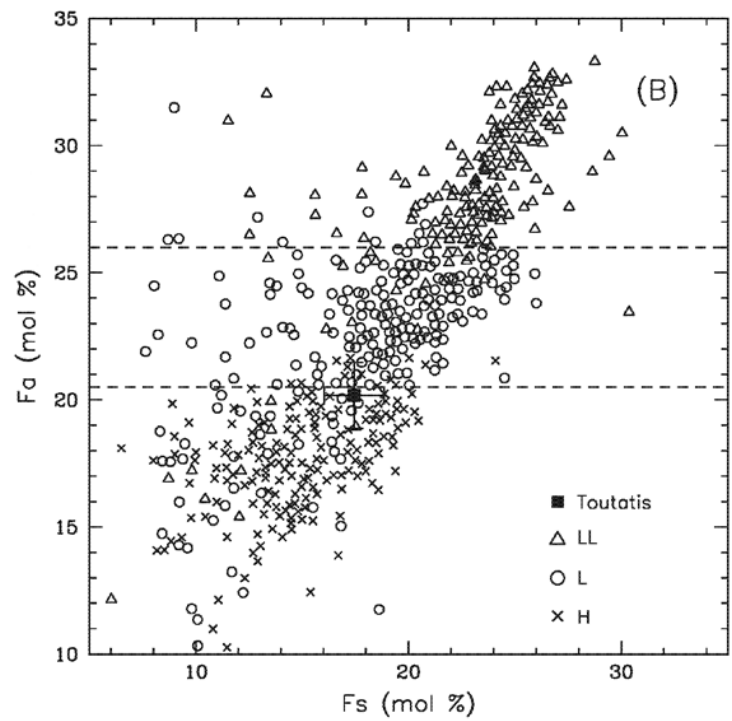

332
333